\documentclass[letterpaper,twocolumn,10pt]{article}
\usepackage{beramono}
\usepackage{listings}
\usepackage{usenix-2020-09}
\usepackage{tcolorbox}
\usepackage{soul}
\usepackage{balance}
\usepackage{listings}
\newcommand{\mybox}[1]{%
	\setbox0=\hbox{#1}%
	\setlength{\@tempdima}{\dimexpr\wd0+13pt}%
	\begin{tcolorbox}[boxrule=0.5pt, colback=white, arc=4pt,
		left=6pt,right=6pt,top=6pt,bottom=6pt,boxsep=0pt]
		#1
	\end{tcolorbox}
}

\definecolor{codegreen}{rgb}{0,0.6,0}
\definecolor{codegray}{rgb}{0.5,0.5,0.5}
\definecolor{codepurple}{rgb}{0.58,0,0.82}
\definecolor{backcolour}{rgb}{0.95,0.95,0.92}

\lstdefinestyle{mystyle}{
  backgroundcolor=\color{backcolour}, commentstyle=\color{codegreen},
  keywordstyle=\color{magenta},
  numberstyle=\tiny\color{codegray},
  stringstyle=\color{codepurple},
  basicstyle=\ttfamily\footnotesize,
  breakatwhitespace=false,         
  breaklines=true,                 
  captionpos=b,                    
  keepspaces=true,                 
  numbers=left,                    
  numbersep=5pt,                  
  showspaces=false,                
  showstringspaces=false,
  showtabs=false,                  
  tabsize=2
}

\definecolor{songcolor}{RGB}{191,191,191}

\newcommand{\tool}{{DeepMut}}

\usepackage{tikz}
\usepackage{amsmath}
\usepackage{tablefootnote}
\usepackage{longtable}
\usepackage{adjustbox}
\usepackage{xcolor}
\usepackage{fvextra}
\usepackage{subcaption}
\usepackage{tcolorbox}
\usepackage[flushleft]{threeparttable}
\usepackage{pgfplots}
\usepackage{pgfplotstable}
\pgfplotsset{compat=1.15}
\usepackage{pictex}
\usepackage{pgfkeys}
\usepackage{tikz}
\usepackage{caption}
\usepackage{graphicx}
\usetikzlibrary{trees,positioning,shapes,shadows,arrows}
\usepackage{multirow}
\makeatletter
\newbox\image@box%
\newdimen\image@width%
\newcommand\IncludeGraphics[2][\@empty]{%
  \setbox\image@box=\hbox{\includegraphics[#1]{#2}}%
  \image@width\wd\image@box%
  \ifdim \image@width>\linewidth%
    \setbox\image@box=\hbox{\includegraphics[width=\linewidth]{#2}}%
    \box\image@box%
  \else%
    \includegraphics[#1]{#2}%
  \fi%
}
\definecolor{mGreen}{rgb}{0,0.6,0}
\definecolor{mGray}{rgb}{0.5,0.5,0.5}
\definecolor{mPurple}{rgb}{0.58,0,0.82}
\definecolor{backgroundColour}{rgb}{0.95,0.95,0.92}

\lstdefinestyle{CStyle}{
    backgroundcolor=\color{backgroundColour},   
    commentstyle=\color{mGreen},
    keywordstyle=\color{magenta},
    numberstyle=\tiny\color{mGray},
    stringstyle=\color{mPurple},
    basicstyle=\footnotesize,
    breakatwhitespace=false,         
    breaklines=true,                 
    captionpos=b,                    
    keepspaces=true,                 
    numbers=left,                    
    numbersep=5pt,                  
    showspaces=false,                
    showstringspaces=false,
    showtabs=false,                  
    tabsize=2,
    language=C
}

\begin{document}

\date{}

\title{\Large \bf Characterizing and Understanding Software Security Vulnerabilities in Machine Learning Libraries}
\author{
{\rm Nima Shiri Harzevili}\\
York University \\
nshiri@yorku.ca
\and
{\rm Jiho Shin}\\
York University\\
jihoshin@yorku.ca
 \and
{\rm Junjie Wang}\\
Chinese Academy of Sciences\\
junjie@iscas.ac.cn
 \and
 {\rm Song Wang}\\
York University\\
wangsong@yorku.ca
} 

\maketitle

\begin{abstract}
The application of machine learning (ML) libraries has been tremendously increased in many domains, including autonomous driving systems, medical, and critical industries. Vulnerabilities of such libraries could result in irreparable consequences. However, the characteristics of software security vulnerabilities have not been well studied. In this paper, to bridge this gap, we take the first step towards characterizing and understanding the security vulnerabilities of five well-known ML libraries, including TensorFlow, PyTorch, Scikit-learn, Pandas, and Numpy. To do so, we collected 596 security vulnerabilities  
to explore five major factors: 1) vulnerability types, 2) root causes, 3) symptoms, 4) fixing patterns, and 5) fixing efforts of security vulnerabilities in ML libraries. The findings of this study can help developers understand the characteristics  of security vulnerabilities across different ML libraries. 
To make our finding actionable, we further developed {\tool}, an automated mutation testing tool, as a proof-of-concept application of our findings. {\tool} is designed to assess the adequacy of existing test suites of ML libraries against security-aware mutation operators extracted from the vulnerabilities studied in this work. We applied {\tool} on the TensorFlow kernel module and found more than 1k alive mutants not covered by the existing test suits, which have been confirmed by the development team of TensorFlow. The results demonstrate the usefulness of our findings.

\end{abstract}

\section{Introduction}
\label{sec:intro}

Nowadays, machine learning (ML) libraries have been frequently used in a wide variety of domains including but not limited to image classification \cite{algan2021image, mahdisoltani2018fine}, big data analysis \cite{patgiri2018taxonomy}, pattern recognition \cite{lv2022semi}, 
self-driving \cite{simhambhatla2019self, ramos2017detecting, kulkarni2018traffic} and Natural Language Processing \cite{minaee2017automatic, athreya2021template, roy2021deep}. These ML libraries can be vulnerable to many attacks~\cite{madry2017towards}, and failures to detect the vulnerabilities in these libraries could cause catastrophic outcomes, such as car accidents~\cite{hong2020artificial}.

In the past years, there have been multiple research studies to characterize ML bugs from end-users' context in which bugs are mainly related to API usage of ML libraries~\cite{humbatova2020taxonomy, islam2019comprehensive, zhang2018empirical, shen2021comprehensive}, or developers' context where the bugs are located inside components or core algorithms of ML libraries (implementation bugs)~\cite{jia2021symptoms, thung2012empirical, garcia2020comprehensive, shen2021comprehensive, di2017comprehensive}. For example, Zhang et al.~\cite{zhang2018empirical} focused on one typical deep learning (DL) library, i.e., TensorFlow, and studied DL application bugs built on top of TensorFlow. 
Islam et al.~\cite{di2017comprehensive} conducted the first study on characterizing API usage bugs of five DL libraries, including Caffe, Keras, TensorFlow, Theano, and Torch. They provided a classification for bug types, root causes, impact, and the DL development stage where bugs occur. 
Despite these efforts, the characteristics of software security vulnerabilities in ML libraries have not been well studied, which leaves unanswered the more directly relevant questions:

\textit{What kinds of security vulnerabilities are
found in ML libraries}? \textit{What are the root causes of security vulnerabilities in ML libraries}? \textit{What symptoms do these security vulnerabilities have}? \textit{Are there any fixing patterns for resolving these security vulnerabilities}? And \textit{What are the efforts required to fix these security vulnerabilities}? 

Understanding such characteristics of security vulnerabilities in ML libraries has the potential to foster the development of secure and reliable ML platforms. 

To fill the above research gap, we take the first step towards characterizing and understanding security vulnerabilities in ML libraries. More specifically, we conduct the first comprehensive study to explore five significant factors: 1) vulnerability types, 2) root causes, 3) symptoms, 4) fixing patterns, and 5) fixing effort  
of security vulnerabilities 
in 
five well-known ML libraries including TensorFlow \cite{abadi2016TensorFlow}, Keras \cite{chollet2015keras}, PyTorch \cite{paszke2019pytorch}, Scikit-learn \cite{scikit-learn}, Pandas \cite{mckinney2010data}, and Numpy \cite{harris2020array}. 
For our study, we consider all available commits when we conduct this study on Sept. 1st, 2021, to collect security vulnerabilities in each ML library.  
We first searched commit messages with keywords that are related to vulnerabilities (details are in Section~\ref{datacoll}) to identify commits that fixed security vulnerabilities. As a result, 4K commits are collected. We then manually check each commit collected in the first step and identify and characterize vulnerabilities from it by following systematic processes (details are in Section~\ref{labelingProcess}).  In total, we obtained 596 unique security vulnerabilities from the studied five ML libraries. In this paper, we are to address the following research questions: 

\textbf{RQ1: What types of vulnerabilities exist in ML libraries?}
During the development and maintenance of ML libraries, developers often have to deal with various vulnerabilities of different types. To better understand vulnerabilities, this research question categorizes vulnerabilities and demonstrates their frequencies and distributions for each library. In this paper, we categorize vulnerability types based on Common Weakness Enumeration (CWE)\footnote{https://cwe.mitre.org/}.

\textbf{RQ2: What are the root causes for vulnerabilities in ML libraries?}
To understand the nature of ML vulnerabilities, it is critical to identify the root cause, which helps developers explore potential approaches to avoiding and fixing vulnerabilities. This research question examines the detailed root causes of vulnerabilities among the studied ML libraries.

\textbf{RQ3: What are the symptoms of vulnerabilities in ML libraries?}
This research question categorizes the symptoms or effects of different vulnerabilities in ML libraries as understanding the symptoms can help developers assess the impact of vulnerabilities and triage them appropriately. In addition, the symptoms can help developers identify vulnerabilities quickly during software testing.

\textbf{RQ4: What are the fixing patterns for vulnerabilities in ML libraries?}
Fixing patterns provide general solutions for resolving specific types of vulnerabilities. In this research question, we study the patches of each vulnerability to identify and analyze its fixing resolution. Common fixing resolutions across multiple vulnerabilities are grouped into different fixing patterns.   

\textbf{RQ5: What are the efforts required for fixing vulnerabilities in ML libraries?} 
Fixing effort can help measure how much effort the development team has allocated to fix vulnerabilities. In this paper we are following existing studies~\cite{di2017comprehensive, tan2014bug,li2017large}  
and use the line of code changed to measure the fixing effort of vulnerabilities.

This paper makes the following contributions: 
\begin{itemize}

\item{To the best of our knowledge, we conduct the first empirical study to  characterize and understand software security vulnerabilities in ML libraries.}

\item {We provide a set of practical guidelines to help machine learning development teams to develop reliable and secure ML libraries.}

\item We develop {\tool}, an automated mutation testing tool for ML libraries as a proof-of-concept application of our findings.  {\tool} is developed to evaluate the adequacy of the existing test suite of ML libraries against security-aware mutation operators extracted from the studied vulnerabilities in ML libraries.

\item We have applied {\tool} on the TensorFlow kernel module and found more than 1k alive mutants that are not covered by the existing test suite of TensorFlow.

\item{We release the dataset and source code of our experiments to help other researchers replicate and extend our study\footnote{https://cse19922021.github.io/Deep-Learning-Security-Vulnerabilities/}.}
\end{itemize}

\section{Methodology}
\label{sec:approach}

\subsection{Data Collection}
\label{datacoll}

This paper studies five widely used ML libraries, including TensorFlow, PyTorch, Scikit-learn, Pandas, and Numpy. The reason is that the studied ML libraries cover the current industrial machine learning practice and represent the critical aspects of machine learning developments. For example, TensorFlow is low-level, while PyTorch provides high-level APIs to hide the low-level details. Scikit-learn is a machine learning library with hundreds of APIs to build various machine learning models. Pandas and Numpy are two famous data analysis and visualization tools focusing on working with arrays and data frames.
Thus, studying these libraries can help provide a comprehensive understanding of software vulnerabilities in ML libraries and further assist practitioners to build more reliable and secure ML libraries. We excluded some popular deep learning libraries, such as Caffe,  Keras, and Theano, from our experiment subjects. The reason is that we cannot get sufficient historical security vulnerabilities from their repositories, i.e., we could only get 15 security-related commits from Keras's GitHub repository. For each library, we consider all available commits when we conduct this study on Sept. 1st, 2021, to collect security vulnerabilities. Table~\ref{VulChar} shows the statistics of our experiment projects. Our vulnerability collection process consists of the following two steps.

\textbf{Step 1}: \textit{Vulnerability fixing commit collection}. We first extracted all the public CVEs of each experimental project available in the National Vulnerability Database (NVD) on Sept. 1st, 2021. We consider commits whose commit messages contain these CVEs as the fixing commits to these vulnerabilities by following existing studies~\cite{song2021,tian2012identifying}. Note that, as reported in existing studies~\cite{wijayasekara2012mining,ponta2019manually}, not all security vulnerabilities have CVE identifiers. For example, in our data collection process, we found that four of the five experimental subjects (i.e., PyTorch, Scikit-Learn, Pandas, Numpy) do not have CVEs so far. To cover all possible vulnerabilities, we used the heuristical approaches proposed by Zhou et al.~\cite{zhou2017automated}, to identify the security fixing commits. Specifically, we follow their study and use their designed regular expression rules, including possible expressions and keywords related to security issues, to collect security vulnerability fixing commits. As a result, we collected 4,152 fixing commits on the five studied projects (details are given in Table \ref{VulChar}).

\textbf{Step 2}: \textit{Vulnerability Identification}. Our heuristic approaches to vulnerability fixing commit collection might contain noise as the approach can introduce false positives~\cite{zhou2017automated}. In addition, an existing study showed that one security vulnerability could have multiple fixing commits, which need to be grouped for having a complete picture of the involved vulnerability~\cite{song2021}. Thus, to reduce noise and make our dataset more accurate, we further conduct a manual analysis to group possible fixing commits and identify unique security vulnerabilities on the data collected in step 1. In particular, the authors separately inspected each candidate vulnerability fixing commits to identify vulnerabilities, and they investigated inconsistent cases together to reach a consensus. Finally, we identified 596 unique vulnerabilities from the 4,152 fixing commits collected in Step 1. 


\begin{table}[t!]
\caption{Statistics of experiment subjects in this study}
\centering
\setlength\tabcolsep{3.0pt}
\renewcommand{\arraystretch}{1}
\resizebox{\columnwidth}{!}{%
\begin{tabular}{lcccc}
\hline
\textbf{ML libraries}   &\textbf{\#CVEs} &\textbf{\#commits} &\textbf{\# Vulnerability} & \textbf{Language} \\
\hline
Tensorflow   & 36 & 1,197             & 250              & C++/Python \\
PyTorch      & N.A & 563              & 75               & C++/Python \\
Sickit-Learn & N.A& 325               & 37               & Python \\
Pandas       & N.A& 869               & 84               & Python \\
Numpy        & N.A& 1,198             & 150              & C/Python \\ \hline
\textbf{Overall}      &  36& 4,152             & 596              & -  \\
\hline
\end{tabular}}
\label{VulChar}
\end{table}


\subsection{Data Labeling}
\label{labelingProcess}
In our study, we analyzed each vulnerability from multiple aspects: 1) vulnerability type, 2) root cause, 3) symptom, 4) fixing pattern, and 5) fixing effort. Please note that some existing studies on analyzing general software bugs in machine learning libraries have also provided taxonomies for these aspects~\cite{islam2019comprehensive,shen2021comprehensive}. In this work, we did not adopt corresponding taxonomies from these studies. The reason is that existing studies merely focus on general software bug characteristics of ML libraries either from end-user or developers' perspectives. In other words, they do not provide categorizations for vulnerabilities of ML libraries. As a result, it is not valid to adopt their classifications since general software bugs' characteristics and security vulnerabilities can be significantly different. 

In our labeling process, two authors work together to review 
each identified vulnerability. In particular, they check the related artifact of this vulnerability, including fixing commits, developer discussion, and pull requests to carefully understand the vulnerability and provide the following information for each vulnerability: 1) vulnerability type, 2) root cause, 3) symptom, 4) fixing pattern, and 5) fixing effort. The two authors discussed the disagreements together during the labeling process until all information was extracted consistently.

\section{Result Analysis}
In this section, we present and discuss our analysis results to address the five research questions we asked in Section~\ref{sec:intro}.

\subsection{RQ1: Vulnerability Types}

\begin{figure*}
     \centering
         \includegraphics[width=\textwidth]{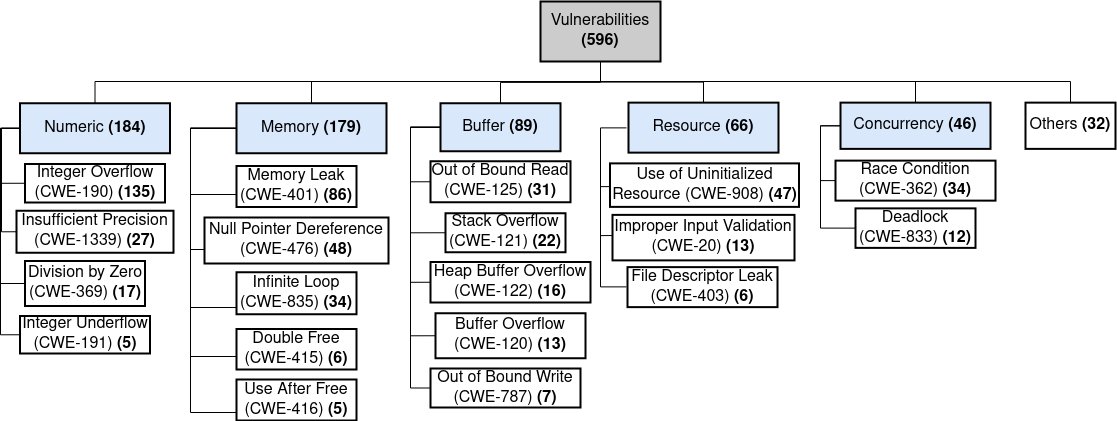}
         \caption{The taxonomy of vulnerability types studied in this work.}
          \label{vuldiag}
\end{figure*}

The taxonomy of vulnerability types is shown in Figure~\ref{vuldiag}. It is organized into five high-level categories (i.e., \textbf{Memory}, \textbf{Resource}, \textbf{Numeric}, \textbf{Buffer}, \textbf{Concurrency}) and involves more than 19 different CWEs covered by 567 (94.6\%) of the 596 vulnerabilities. The remaining 32 (5.34\%) vulnerabilities appear infrequently and do not belong to any particular vulnerabilities, and are included in \textbf{Others} category.

\noindent \textbf{Numeric.} 
Vulnerabilities in this category mostly deal with improper calculation or conversion of numbers, accounting for 184 (30.8\%) of the vulnerabilities. It mainly has four types of CWEs: 1) \textit{Integer Overflow (CWE-190)}, 2) \textit{Insufficient Precision or Accuracy of a Real Number (CWE-1339)}, 3) \textit{Division by Zero (CWE-369)}, and 4) \textit{Integer Underflow (CWE-191)}.

\noindent \textbf{Memory.} This category contains vulnerabilities consuming GPU memory abnormality, accounting for 179 (30\%) of the vulnerabilities. Specifically, it contains the following five types of CWEs: 1) \textit{Missing Release of Memory after Effective Lifetime (Memory Leak (CWE-401)}, 2) \textit{Null Pointer Dereference (CWE-476)}, 
3) \textit{Infinite Loop (CWE-835)}, 4) \textit{Double Free (CWE-415)} , and 5) \textit{Use After Free (CWE-416)}.

\noindent \textbf{Buffer. }
This type of vulnerability corresponds to the handling of memory buffers within a software system, accounting for 89 (14.9\%) of the vulnerabilities. It mainly covers five types of CWEs: 1) \textit{Out of Bound Read (CWE-125)}, 2) \textit{Stack Overflow (CWE-121)}, 3) \textit{Heap Buffer Overflow (CWE-122)}, 4) \textit{Buffer Overflow (CWE-120)}, and 5)\textit{Out of Bound Write (CWE-787)}.

\noindent \textbf{Resource.} 
Vulnerabilities in this category correspond to resource initialization or validation issues, accounting for 66 (11\%) of the vulnerabilities. It consists of three types of CWEs: 1) \textit{Use of Uninitialized Resource(CWE-908)}, 2) \textit{Improper Input Validation (CWE-20)}, and 3) \textit{File Descriptor Leak (CWE-403)}.

\noindent \textbf{Concurrency.} 
Vulnerabilities in this category relate to concurrent access of resources and their locking applied by multiple threads, accounting for 46 (11.5\%) vulnerabilities. It consists of two types of CWEs: 1) \textit{Race Condition (CWE-362)}
and 2) \textit{Deadlock (CWE-833)}.

\definecolor{ForestGreen}{RGB}{34,139,34}

\pgfplotstableread{
Label Tensorflow PyTorch Sickit-Learn Numpy Pandas
NumericErrors 70 27 22 24 41
MemoryErrors 52 16 10 86 15
BufferErrors 49 12 1 15 12
ResourceErrors 27 13 0 13 13
ConcurrencyErrors 39 2 1 4 0
Others 13 5 3 8 3
}\testdata
    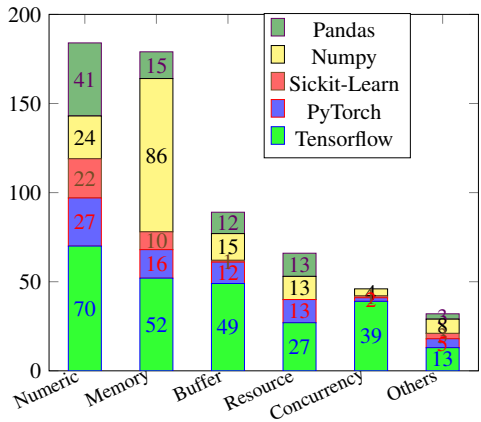
\begin{figure}[t!]
     \centering
  \renewcommand{\arraystretch}{1}
\resizebox{0.75\columnwidth}{!}{%
    \begin{tikzpicture}[scale=0.4]
    \begin{axis}[
        ybar stacked,
    	bar width=15pt,
        ymin=0,
        ymax=200,
        xtick=data,
        legend style={at={(0.5,0.8)},anchor=west},
        reverse legend=true,
        xticklabels={\small{Numeric}, \small{Memory}, \small{Buffer}, \small{Resource}, \small{Concurrency}, \small{Others}},
        xticklabel style={rotate=25,anchor=east},
    ]
    \addplot + [ nodes near coords, fill=green!80] table [y=Tensorflow, meta=Label, x expr=\coordindex] {\testdata};
    \addplot +[nodes near coords, fill=blue!60] table  [y=PyTorch, meta=Label, x expr=\coordindex] {\testdata};
    \addplot + [nodes near coords, fill=red!60] table   [y=Sickit-Learn, meta=Label, x expr=\coordindex] {\testdata};
    \addplot + [nodes near coords, fill=yellow!60] table  [y=Numpy, meta=Label, x expr=\coordindex] {\testdata};
    \addplot + [nodes near coords, fill=ForestGreen!60] table  [y=Pandas, meta=Label, x expr=\coordindex] {\testdata};
\legend{Tensorflow, 
PyTorch, 
Sickit-Learn,
Numpy,
Pandas 
}
\end{axis}
\end{tikzpicture}}
\caption{The distribution of software vulnerabilities in different ML libraries.}
\label{typeDist}
\end{figure}

Figure \ref{typeDist} shows the distribution of each type of vulnerability in the studied five libraries. As we can see, \textbf{Numeric} and \textbf{Memory} are the two dominating types across all libraries. Specifically, \textbf{Numeric} is the most common type of vulnerability among libraries except Numpy library, where \textit{Memory Leak} is the most common vulnerability. We demonstrate distribution of subcategories for \textbf{Numeric} and \textbf{Memory} in Table \ref{tableNumericErrors} and Table \ref{tableMemoryBufferErrors}. As shown in Table \ref{tableNumericErrors}, ~\textit{Integer Overflow (CWE-190)} is the most common vulnerability type among other subcategories for all libraries. This states that \textit{Integer Overflow} in the studied ML libraries is critical, and developers need to pay more attention to this type.

\begin{tcolorbox}[width=8.5cm]
\textbf{Finding 1:} Vulnerabilities in the categories of \textbf{Numeric} and \textbf{Memory} are most frequent vulnerabilities across all libraries, and \textit{Integer Overflow} is the most common vulnerability in the five ML libraries. 
\end{tcolorbox}

\begin{table}[t!]
\caption{Subcategories of \textbf{Numeric}.}
\label{tableNumericErrors}
\centering
\setlength{\tabcolsep}{5pt} 
\renewcommand{\arraystretch}{1}
\resizebox{\columnwidth}{!}{%
\begin{tabular}{ccccc}
\hline
 Library& \small{Integer Overflow} & \small{Insufficient Precision}  &  \small{Division by Zero}   &  \small{Integer Underflow}\\
                         \hline
Tensorflow               & 63               &    4              &  2                             & 1                \\
PyTorch                  & 18               &    6              &   3                            & 0                \\
Sickit-learn             & 8                &     4              &    7                           & 3                \\
Pandas                   & 34               &    3              &    3                             & 1                \\
Numpy                    & 12               &    10              &    2                           & 0                \\
\hline \hline
\textbf{Sum}                      & 135              & 27                & 17                     & 5  \\   \hline
\end{tabular}
}
\end{table}
\begin{table}[t!]
\centering
\caption{Subcategories of \textbf{Memory}.}
\label{tableMemoryBufferErrors}
\setlength{\tabcolsep}{5pt} 
\renewcommand{\arraystretch}{1}
\resizebox{\columnwidth}{!}{%
\begin{tabular}{cccccc}
\hline
Library& \small{Memory Leak} & \small{Null Pointer Dereference} & \small{Infinite Loop} & \small{Double Free} & \small{Use After Free} \\
\hline
Tensorflow               & 6           & 25                       & 19            & 1           & 1              \\
PyTorch                  & 2           & 7                        & 5             & 2           & 0              \\
Sickit-learn             & 8           & 0                        & 2             & 0           & 0              \\
Pandas                   & 5           & 6                        & 3             & 1           & 0              \\
Numpy                    & 65          & 10                       & 5             & 2           & 4              \\ \hline \hline
\textbf{Sum}                      & 86          & 48                       & 34            & 6           & 5            \\
\hline
\end{tabular}
}
\end{table}


\begin{center}
\begin{figure*}[t]
    \centering
     \includegraphics[width=16cm]{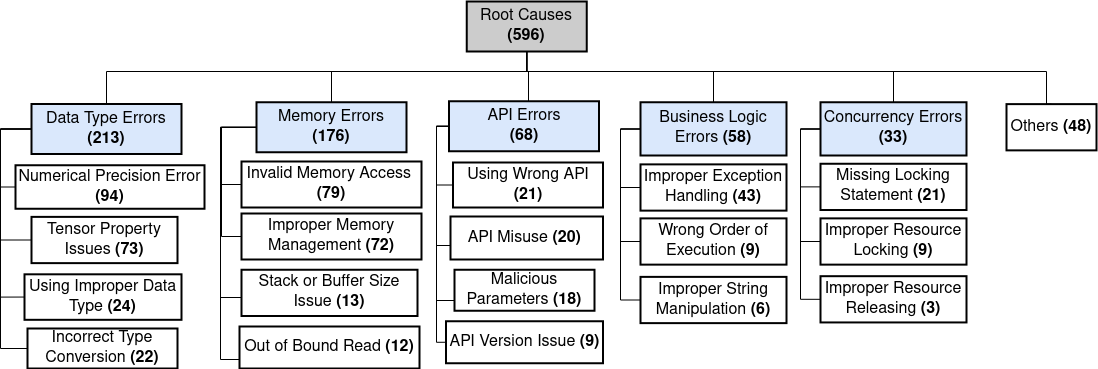}
    \caption{Taxonomy of root causes in ML libraries.} 
    
    \label{rootcausediag}
\end{figure*}
\end{center}
\subsection{RQ2: Root Causes}
\label{section:RQ2}

The taxonomy of root causes of studied vulnerabilities in ML libraries is shown in Figure~\ref{rootcausediag}, which is organized into five high-level categories including \textbf{Data Type Errors}, \textbf{Memory Errors}, \textbf{API Errors}, \textbf{Business Logic Errors}, and \textbf{Concurrency Errors} covered by 558 (91.9\%) of the 596 vulnerabilities. The remaining 48 (5.1\%) root causes have no clear indication about their types, and hence we group them in \textbf{Others} category.

\noindent \textbf{Data Type Errors.}
Root causes in this category mostly deal with range or precision issues of conventional data types defined by developers accounting for 213 (35.7\%) of vulnerabilities. Subcategories include 1) \textit{Numerical Precision Error}: When developers define variables or tensors with a limited or large range, 2) \textit{Tensor Property Issue}: When a thread or a program maintains tensors inappropriately, 3) \textit{Using Improper Data Type}: When developers have confusion about using data types, e.g., using int32 instead of uint32, 4) \textit{Incorrect Type Conversion}: When a developer incorrectly convert data types together, e.g., implicit type conversion of float to double. 

\noindent \textbf{Memory Errors.} 
Root causes in this category mainly deal with memory-related vulnerabilities accounting for 29.5\% of vulnerabilities. The subcategories of this root cause are including 1) \textit{Invalid Memory Access:} When processes try to access memory locations filled with null values and already been deleted or freed,
2) \textit{Improper Memory Management:} When a developer has confusion in memory management, either misuse memory release statement or forget to release memory after its lifetime,
3) \textit{Out of Bound Read:} When processes read information from other memory locations, and 4) \textit{Stack or Buffer Size Issue:} When developers define the stacks or buffers with inappropriate sizes.  

\noindent \textbf{API Errors.}
This root cause category is due to inconsistencies in updating or using APIs accounting for 58 (11.4\%) of total records. Subcategories include 1) \textit{API Misuse}: When developers mistakenly use a specific API, e.g., passing parameters in wrong orders, lack of using optional parameters, mistakenly using optional parameters, etc., 2) \textit{Using Wrong API}: When developers mistakenly use improper APIs, 3) \textit{Malicious Parameters}: When developers pass malicious or invalid parameters to API calls which are exploitable by attackers. Attackers can exploit these parameters by crafting particular inputs to take control of the software system, and 4) \textit{API Version Issue}: When developers mistakenly use either wrong versions of APIs or outdated ones.

\noindent \textbf{Business Logic Errors.} This root cause accounts for 58 (9.7\%) of vulnerabilities. It includes 1) \textit{Improper Exception Handling}: When developers incorrectly handle exceptional conditions leading to termination of the software during normal executions of the software, 2) \textit{Wrong Order of Execution}: When a set of steps or components were inappropriately executed, 3) \textit{Improper String Manipulation}: When developers parse string variables or absolute and relative addresses incorrectly.

\noindent \textbf{Concurrency Errors.}
This root cause category involves concurrent access of resources in a shared environment by multiple threads due to improper resource locking, releasing, or simultaneous resource access accounting for 33 (5.5\%) vulnerabilities. Subcategories are 1) \textit{Missing Locking Statement}: When developers forget to lock resources which mostly result in race condition or deadlock errors, 2) \textit{Improper Resource Locking}: When developers use locking statements improperly on program resources, 3) \textit{Improper Resource Releasing}: When developers release locked resource inappropriately, which can result in deadlock or race condition errors. 



\pgfplotstableread{
Label Tensorflow PyTorch Sickit-Learn Numpy Pandas
DataTypeErrors 100 35 14 28 36
MemoryErrors 57 19 7 78 15
APIErrors 14 5 8 26 15
InconsistentBusinessLogic 27 9 8 8 6
Others 25 4 0 7 12
ConcurrencyErrors 27 3 0 3 0
    }\testdata
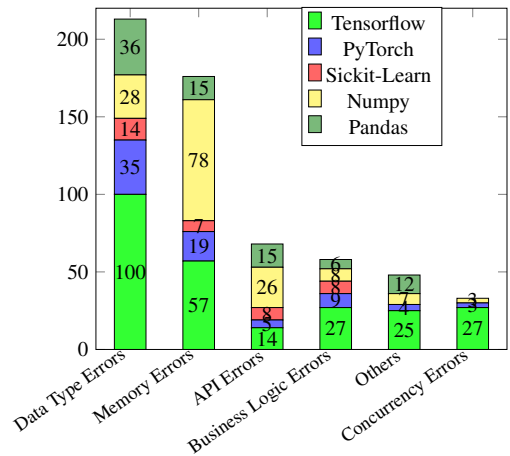
\begin{figure}[t!]
     \centering
  \renewcommand{\arraystretch}{1}
\resizebox{0.8\columnwidth}{!}{%
    \begin{tikzpicture}
    \begin{axis}[
        ybar stacked,
    	bar width=15pt,
        ymin=0,
        ymax=220,
        xtick=data,
        legend style={at={(0.5,0.8)},anchor=west},
        xticklabels={\small{Data Type Errors}, \small{Memory Errors},  \small{API Errors}, \small{Business Logic Errors}, \small{Others}, \small{Concurrency Errors}},
        xticklabel style={rotate=35,anchor=east},
    ]
    \addplot  [ nodes near coords, fill=green!80] table [y=Tensorflow, meta=Label, x expr=\coordindex] {\testdata};
    \addplot [nodes near coords, fill=blue!60] table  [y=PyTorch, meta=Label, x expr=\coordindex] {\testdata};
    \addplot [nodes near coords, fill=red!60] table   [y=Sickit-Learn, meta=Label, x expr=\coordindex] {\testdata};
    \addplot [nodes near coords, fill=yellow!60] table  [y=Numpy, meta=Label, x expr=\coordindex] {\testdata};
    \addplot [nodes near coords, fill=ForestGreen!60] table  [y=Pandas, meta=Label, x expr=\coordindex] {\testdata};
\legend{Tensorflow, 
PyTorch, 
Sickit-Learn,
Numpy,
Pandas, 
}
\end{axis}
\end{tikzpicture}
}
\vspace{-0.1in}
  \caption{Distribution of root causes across libraries}
  \label{rootDist}
\end{figure}

Figure~\ref{rootDist} shows the distribution of root causes of vulnerabilities in different libraries. As can be seen, \textbf{Data Type Errors} is the most common root cause of vulnerabilities across the studied ML libraries. Table~\ref{dterror} further shows the distribution of subcategories of~\textbf{Data Type Errors} across the studied ML libraries. As we can see, \textit{Numerical Precision Error} is the major subcategory because the studied ML libraries mostly rely on tensor level and array level computations. The computations involve quantization during training which means millions of parameters are multiplying or adding together in integer or float types to make models as smaller as possible. \textit{Tensor Property Issue} is the second most common root cause of vulnerabilities. The reason is that optimization and quantization operations are mostly done with tensors.

\begin{table}[t!]
\centering
\caption{Subcategories of \textbf{Data Type Errors}}
\label{dterror}
\renewcommand{\arraystretch}{1}
{%
\begin{tabular}{lllll}
\hline
Library& NPE$^1$& TPI$^2$ & UDP$^3$ & ITC$^4$  \\
\hline
Tensorflow               & 34      &   47               & 13                       & 6 \\
PyTorch                  & 17       &   11              & 5                        & 2  \\
Sickit-learn             & 12       &     0            & 1                        & 1   \\
Pandas                   & 21       &      2           & 4                        & 9    \\
Numpy                    & 10       &      13           & 1                        & 4     \\ \hline \hline
\textbf{Sum}             & 94       &   73           & 24                       & 22     \\                  
\hline
\end{tabular}}
\vskip3pt
{\small $^1$ Numerical Precision Error $|$
$^2$ Tensor Property Issues $|$
$^3$ Using Improper Data Type $|$
$^4$ Incorrect Type Conversion}
\end{table}

\textbf{Memory Errors} is the second most common root cause of vulnerabilities across the studied ML libraries. Table~\ref{memoryerror} further shows the distribution of \textbf{Memory Errors}. As you can see, \textbf{Invalid Memory Access} is the major root cause of memory-related vulnerabilities. An invalid memory is a memory that is undefined, uninitialized, deleted, containing null values, corrupted values, erased, etc, \textbf{Improper Memory Management} is the second most common subcategory of \textbf{Memory Errors}. \textbf{Improper Memory Management} is the major root cause of the Numpy library, which is written in C language and memory management is the responsibility of developers. Developers of Numpy often forget to release the allocated memory address when its effective lifetime is finished.

\begin{tcolorbox}[width=8.5cm]
\textbf{Finding 2:} \textbf{Data Type Errors} and \textbf{Memory Errors} are the most common types of root cause of vulnerabilities accounting for 64.2\% of vulnerabilities in the studied ML libraries respectively. \textit{Numerical Precision Error} is the dominating subcategory.
\end{tcolorbox}

\begin{table}[t!]
\caption{Subcategories of  \textbf{Memory Errors}}
\label{memoryerror}
\setlength{\tabcolsep}{14pt} 
\renewcommand{\arraystretch}{1}
\resizebox{\columnwidth}{!}{%
\begin{tabular}{lllll}
\hline
Library& IMA$^1$ &  IMM$^2$ & SBSI$^3$  &  OOBR$^4$ \\
\hline
Tensorflow   & 41 & 4  & 7  &  5  \\
PyTorch      & 17 & 1  & 1  &  0 \\
Sickit-learn & 0  & 6  & 0  &  1 \\
Pandas       & 8  & 3  & 1  &  3 \\
Numpy        & 13 & 58 & 4  &  3 \\
\hline \hline
\textbf{Sum} & 79 & 72 & 13 & 12 \\
\hline
\end{tabular}

}
{\small $^1$Invalid Memory Access $|$
$^2$Improper Memory Management $|$
$^3$Stack or Buffer Size Issue.
$^4$Out of Bound Read$|$
}
\end{table}

\textbf{API Errors} are the third most common root causes of vulnerabilities in the studied ML libraries. A more detailed distribution of subcategories are shown in Table \ref{apierror}. As you can see, \textbf{Using Wrong API} is the most common subcategory. The second common subcategory is \textbf{API Misuse} where developers have a hard time using APIs, e.g., passing wrong parameters, lack of using optional parameters, and improperly using optional parameters. \textbf{Malicious Parameters} are also common where developers give unsafe inputs to API calls which attackers can exploit. Sometimes developers use outdated or invalid APIs, which are the root cause of vulnerabilities in the studied ML libraries. We categorize them as \textbf{API Version Issue}. 

\textbf{API Errors} differ noticeably compared to traditional software systems. According to a current study on \textbf{API Errors} conducted by Amann et al. \cite{amann2018systematic}, missing and redundant API calls are the most frequent errors. At the same time, these are the least errors in the studied ML libraries. We conclude that developers of the studied ML libraries have difficulty understanding which APIs they should use, how to use them, and make them secure.

\begin{tcolorbox}[width=8.5cm]
\textbf{Finding 3:} \textbf{API Errors} in the studied ML libraries cover more corner cases that are exploitable by attackers compared to traditional software systems. 

\end{tcolorbox}

\begin{table}[t!]
\caption{Subcategories of \textbf{API Errors}}
\centering
\label{apierror}
\renewcommand{\arraystretch}{1}
{%
\begin{tabular}{lcccc}
\hline
Library      & UWA$^1$ & AM$^2$ & MP$^3$ & AVI$3$ \\
\hline
Tensorflow   & 3               & 3          & 3                    & 5                 \\
PyTorch      & 2               & 1          & 1                    & 1                 \\
Sickit-learn & 2               & 3          & 3                    & 0                 \\
Pandas       & 3               & 4          & 7                    & 1                 \\
Numpy        & 11              & 9          & 4                    & 2                 \\
Sum          & 21              & 20         & 18                   & 9               \\
\hline \hline
\textbf{Sum}                      & 21         & 20              & 18                   & 9         \\
\hline
\end{tabular}}

{\small 
$^1$ Using Wrong API $|$
$^2$ API Misuse $|$
$^3$ Malicious Parameters $|$
$^4$ API Version Issue}
\end{table}

\subsection{RQ3: Symptoms}

\begin{figure}[t!]
    \centering
      \includegraphics[width=9cm]{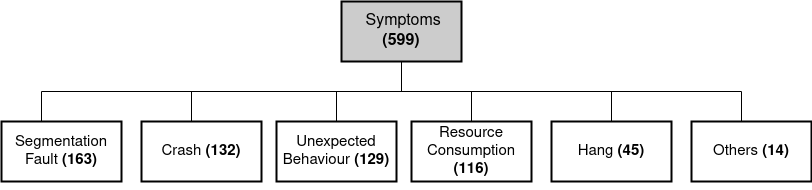}
    \caption{Taxonomy of symptoms in the studied ML libraries}
   \label{symptomsdiag}
\end{figure}

The taxonomy of symptoms of studied vulnerabilities in ML libraries is shown in Figure~\ref{symptomsdiag}, which is organized into six categories including  \textbf{Segmentation Fault}, \textbf{Crash}, and \textbf{Unexpected Behaviour}, \textbf{Resource Consumption}, and \textbf{Hang}, covered by 582 (97.6\%) vulnerabilities. The remaining 14 (2.3\%) symptoms have no clear indication about their outcome, and hence we group them in \textbf{Others} category.

\definecolor{ForestGreen}{RGB}{34,139,34}

\pgfplotstableread{
Label Tensorflow PyTorch Sickit-Learn Numpy Pandas
SegmentationFault 69 30 0 37 26
Crash 73 18 10 16 16
UnexpectedBehavior 47 18 17 18 27
ResourceConsumption 32 2 7 70 5
Hang 29 6 2 5 2
Others 0 1 1 4 8
    }\testdata
    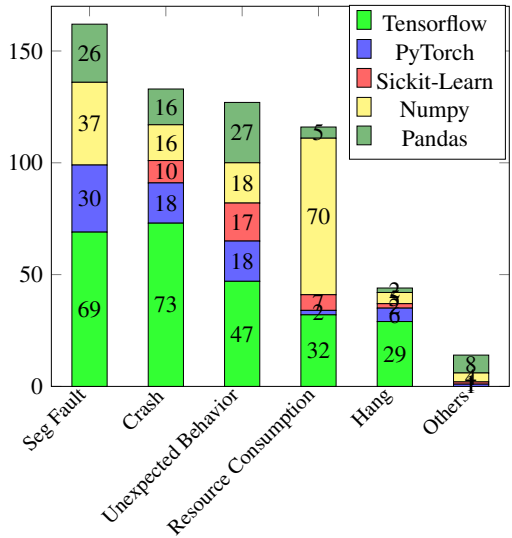
\begin{figure}[t!]
     \centering
  \renewcommand{\arraystretch}{1}
\resizebox{0.8\columnwidth}{!}{%
    \begin{tikzpicture}
    \begin{axis}[
        ybar stacked,
    	bar width=15pt,
        ymin=0,
        ymax=170,
        xtick=data,
        legend style={at={(0.65,0.8)},anchor=west},
        xticklabel style={rotate=45,anchor=east},
        xticklabels={\small{Seg Fault}, \small{Crash}, \small{Unexpected Behavior}, \small{Resource Consumption}, \small{Hang}, \small{Others}},
    ]
    \addplot  [ nodes near coords, fill=green!80] table [y=Tensorflow, meta=Label, x expr=\coordindex] {\testdata};
    \addplot [nodes near coords, fill=blue!60] table  [y=PyTorch, meta=Label, x expr=\coordindex] {\testdata};
    \addplot [nodes near coords, fill=red!60] table   [y=Sickit-Learn, meta=Label, x expr=\coordindex] {\testdata};
    \addplot [nodes near coords, fill=yellow!60] table  [y=Numpy, meta=Label, x expr=\coordindex] {\testdata};
    \addplot [nodes near coords, fill=ForestGreen!60] table  [y=Pandas, meta=Label, x expr=\coordindex] {\testdata};
\legend{Tensorflow, 
PyTorch, 
Sickit-Learn,
Numpy,
Pandas, 
}
\end{axis}
\end{tikzpicture}}
  \caption{Distribution of symptoms across different libraries.}
  \label{sym}
\end{figure}
\begin{figure}[t]
    \centering
     \resizebox{1\columnwidth}{!}{{\includegraphics{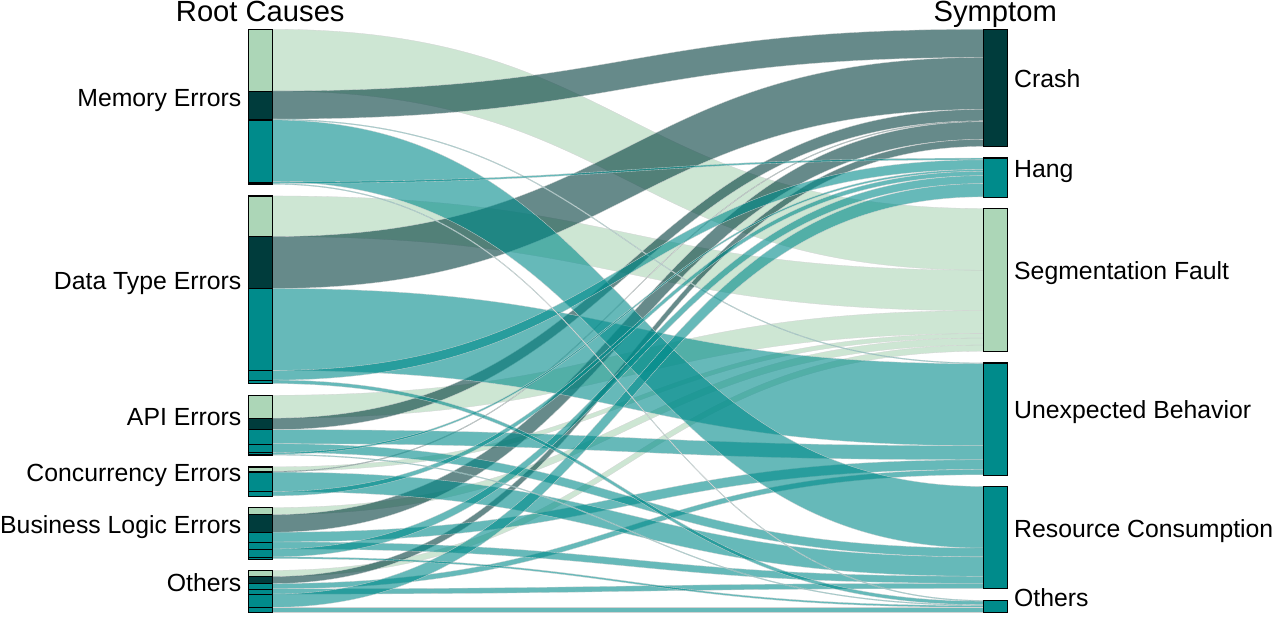}}}\    \caption{Mapping of root causes to symptoms.}
    \label{root2symp}
\end{figure}



\noindent \textbf{Unexpected Behavior}: If the library is producing results or behaving that is not expected. For example, in this \textbf{Integer Overflow} vulnerability from sklearn library\footnote{https://github.com/scikit-learn/\\scikit-learn/commit/622f912095308733ddfe572a619b1574b9da335e} where $pk * qk$ returns \textit{inf} instead of \textit{float} because of exceeding int32 bits limits during multiplication.

Other symptoms includes \textbf{Segmentation Fault}: When a program outputs \textit{core dumped} in the output which is the symptom for segmentation fault\footnote{https://stackoverflow.com/questions/49092527/\\illegal-instructioncore-dumped-tensorflow},  
\textbf{Resource Consumption}: Exhaustion of available resource, e.g., increasing memory usage because of uncontrolled or improper allocation of main memory. \textbf{Hang}: This means that a program keeps running for a long period without termination or responding, and \textbf{Crash}: When a program or process terminates unexpectedly at run-time.

Figure~\ref{sym} demonstrates the distribution of symptoms across different libraries. As you can see, the most frequent symptom is \textbf{Segmentation Fault} accounting for 27.3\% of vulnerabilities. We also draw a mapping from root causes to symptoms to interpret what is the outcome of vulnerabilities as shown in Figure \ref{root2symp}. It is observable from Figure \ref{root2symp} that vulnerabilities caused by \textbf{Memory Errors}, \textbf{Data Type Errors}, and \textbf{API Errors} often have \textbf{Segmentation Fault} as their symptom. Also, we can see the same pattern for \textbf{Crash} symptoms. Often developers can extract descriptive information from \textbf{Segmentation Fault} and \textbf{Crash}. One possible usage scenario of the extracted information is that they can parse stack traces of failed test suits to analyze the exact root causes of the vulnerability and how to locate them. Also, Developers do not need to develop test oracles in order to understand \textbf{Segmentation Fault} and \textbf{Crash} symptoms.

\begin{tcolorbox}[width=8.5cm]
\textbf{Finding 4:} \textbf{Segmentation Fault} and \textbf{Crash} are the most common symptoms of vulnerabilities accounting for 27.3\% and 22.1\% of vulnerabilities respectively. These symptoms can help developers of the studied ML libraries to understand and locate the root cause of vulnerabilities with their descriptive information.
\end{tcolorbox}

\textbf{Unexpected Behaviour} is the third most common symptom of vulnerabilities accounting for 21.3\% of vulnerabilities. As shown in Figure \ref{root2symp}, often \textbf{Unexpected Behaviour} is the symptom of vulnerabilities caused by \textbf{Data Type Errors}. Hence, developers of the studied ML libraries need to equip the existing test suite with test oracles to understand that ML libraries' components are working as they are expected to do so.

\begin{tcolorbox}[width=8.5cm]
\textbf{Finding 5:} \textbf{Unexpected Behaviour} is the third most common symptom of vulnerabilities accounting for 21.3\% of vulnerabilities. Its prevalence might suggest that developers need to develop test oracles to understand that the studied ML libraries meet their requirements defined by either end-users or developers.
\end{tcolorbox}

\begin{center}
\begin{figure*}[t!]
    \centering
     \makebox[\textwidth]{\includegraphics[width=18cm]{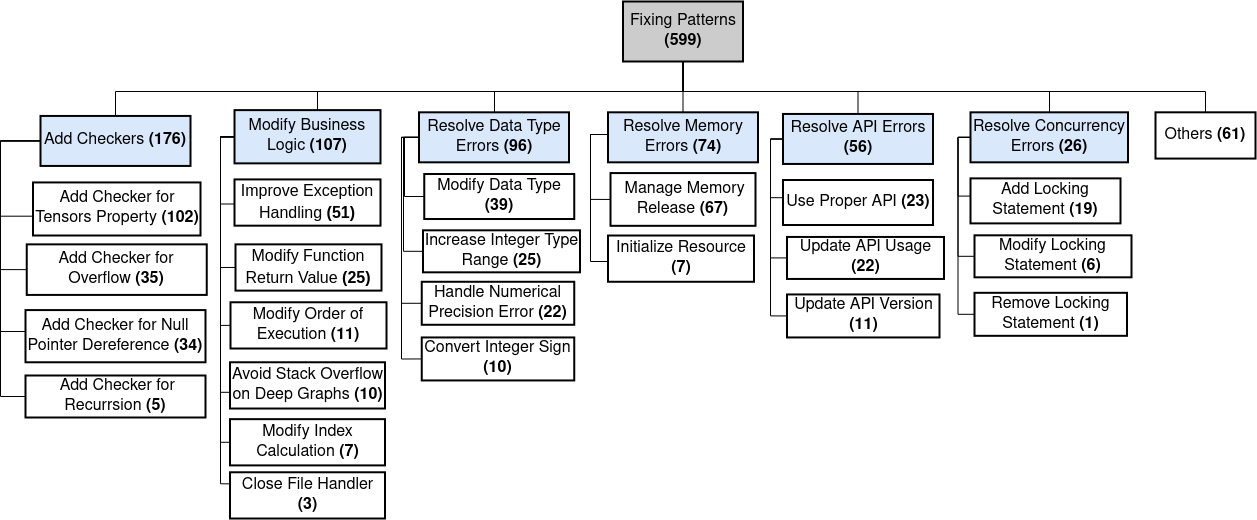}}
    \caption{Taxonomy of fixing patterns in ML libraries.
    }
    \label{fixingdiag}
\end{figure*}
\end{center}

\subsection{RQ4: Fixing Patterns}
The taxonomy of fixing patterns of ML vulnerabilities is shown in Figure \ref{fixingdiag}, which is organized into six high-level categories including \textbf{Add Checkers}, \textbf{Resolve Data Type Errors}, \textbf{Resolve Memory Errors}, \textbf{Resolve API Errors}, \textbf{Resolve Concurrency Errors}, and \textbf{Modify Business Logic Errors} covered by 535 (89.7\%) of the 596 vulnerabilities. The remaining 61 (10.2\%) fixing patterns have no clear indication about their types, and hence are included in the \textbf{Others} category.

\noindent \textbf{Add Checkers.} Fixing patterns in this category are mainly about the addition of either library-specific checkers or conventional checkers to fix vulnerabilities, which cover 176 (29.5 \%) vulnerabilities. Subcategories are 1) \textit{Add Checker for Tensors Property}: This is the most common fixing pattern where developers use if conditions or library-specific checkers to check tensor or arrays properties, e.g., shapes, ranks, values, or elements.

2) \textit{Add Checker for Overflow}: This fixing pattern is mainly used to fix overflow vulnerabilities where developers either add if modules and library-specific checkers or add functions for mathematical operations instead of using explicit mathematical operators, 3) \textit{Adding Checker for Null Pointer Dereference:}  
Developers often add checkers either using if conditions or library-specific checkers to fix null pointer dereferences, and 4) \textit{Adding Checker for Recursion}:
Sometimes an infinite loop occurs due to endless recursion calls. So, developers need to detect recursions that consume stack space, and checkers are added to fix these types of vulnerabilities.

\begin{figure*}[t]
    \centering
     \makebox[\textwidth]{\includegraphics[width=15cm]{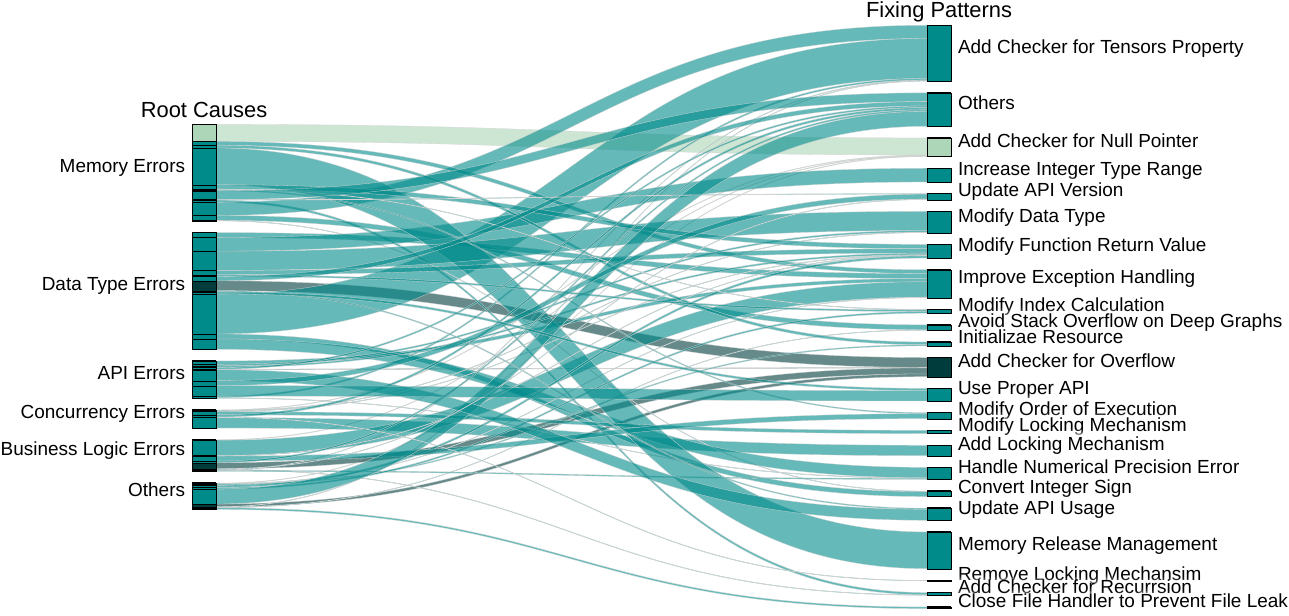}}
    \caption{Mapping of root causes to fixing patterns.}
    \label{rootsymp}
\end{figure*}

\noindent \textbf{Modify Business Logic}:
This type of fixing pattern is related chiefly to modifying the existing control flows, functions, or classes to fix vulnerabilities that have incorrect logic. Subcategories are 1) \textit{Improved Exception Handling}: When the program crashes, it is necessary to record the error or warning message to help developers with debugging. This pattern adds missing error reporting or modifying existing ones that are defective, 2) \textit{Modifying Function Return Value}: This pattern mainly changes function return values to avoid potential mismatches in the data flow of a program for fixing vulnerabilities, 3) \textit{Modify Order of Execution}: Developers change the location of semantically related statements to fix vulnerabilities, 4) \textit{Avoid Stack Overflow on Deep Graphs}: This pattern is used when developers try to prevent stack overflow caused by small stack size or deep computation graphs created in runtime, 5) \textit{Modify Index Calculation}: This pattern is used to fix vulnerabilities related to incorrect indices of data collection such as arrays or tensors, and 6) \textit{Close File Handler to Prevent File Leak}: It is used to fix file descriptor leak vulnerability. 

\noindent \textbf{Resolve Data Type Errors}: Fixing patterns in this category focus on resolving vulnerabilities related to data types, which cover 16.1 \% of the studied ML vulnerabilities. Subcategories include:  1) \textit{Modify Data Type}: is used to fix vulnerabilities involving incorrect data type defined and used, 
2) \textit{Increase Integer Type Range}: is used to fix vulnerabilities that are caused by the limited range of integer types for preventing integer overflow or integer truncation, e.g., using int64 instead of int32, 3) \textit{Handle Numerical Precision}: is used to resolve data type precision issues, e.g., normalization of matrix values during float 16 bits model training, 4) \textit{Convert Integer Sign}: is used to convert data type signs to prevent integer overflow or underflow, e.g., using size\_t instead of int32. 

\noindent \textbf{Resolve Memory Errors.}
Fixing patterns in this category relate to memory management efforts, which can help fix 74 (12.4\%) of total vulnerabilities. Subcategories are 1) \textit{Manage Memory Release}: is used to fix vulnerabilities related to incorrect or inappropriate memory allocations and 2) \textit{Resource Initialization}: when developers initialize tensors, variables, or data types to fix vulnerabilities. 

\noindent \textbf{Resolve API Errors}: Fixing patterns in this category are mainly used to fix vulnerabilities introduced by inappropriate API usages, which help fix 56 (9.3 \%) of vulnerabilities studied in this paper. The detailed subcategories are 1) \textit{Using Proper API}, 2) \textit{Update API Usage}, and 3) \textit{Update API Version}.

\noindent \textbf{Resolve Concurrency Errors.} Fixing patterns in this category are used to fix vulnerabilities related to concurrency issues resulting in deadlock or race condition errors. Subcategories include 1) \textit{Add Locking Mechanism}, 2) \textit{Modify Locking Mechanism}, and  
3) \textit{Remove Locking Mechanism}.

Figure \ref{fixingpatterndist} shows the distribution of fixing patterns across different libraries. As can be seen, \textbf{Adding Checkers} is the most common fixing pattern in ML libraries, accounting for 96, 24, 6, 32, 18 vulnerabilities of TensorFlow, PyTorch, Scikit-Learn, Pandas, and Numpy, respectively. In total, 29.5\% of vulnerabilities can be fixed by this pattern.
We further show the breakdown of \textbf{Adding Checkers} regarding the distributions of subcategories in Table~\ref{checkerstable}. As we can see, \textit{Adding Checker for Tensor Property} is the principal subcategory that covers 102 vulnerabilities.


\definecolor{ForestGreen}{RGB}{34,139,34}

\pgfplotstableread{
Label Tensorflow PyTorch Sickit-Learn Numpy Pandas
AddCheckers 96 24 6 32 18
ModifyBusinessLogic 27 17 12 17 34
ResolveDataTypeErrors 44 18 8 10 16
ResolveMemoryErrors 6 2 4 58 4
Others 44 6 0 7 4
ResolveAPIErrors 13 5 7 23 8
ResolveConcurrencyError 20 3 0 3 0
    }\testdata
    \begin{figure}[t!]
     \centering
  \renewcommand{\arraystretch}{1}
\resizebox{0.8\columnwidth}{!}{%
    \begin{tikzpicture}
    \begin{axis}[
        ybar stacked,
    	bar width=15pt,
        ymin=0,
        ymax=190,
        xtick=data,
        legend style={at={(0.5,0.8)},anchor=west},
        xticklabels={\small{Adding Checkers}, \small{Modify Business Logic},  \small{Resolve Data Type Errors}, \small{Resolve Memory Errors}, \small{Others}, \small{Resolve API Errors}, \small{Resolve Concurrency Error}},
        xticklabel style={rotate=45,anchor=east},
    ]
    \addplot [ nodes near coords, fill=green!80] table [y=Tensorflow, meta=Label, x expr=\coordindex] {\testdata};
    \addplot [nodes near coords, fill=blue!60] table   [y=PyTorch, meta=Label, x expr=\coordindex] {\testdata};
    \addplot [nodes near coords, fill=red!60] table    [y=Sickit-Learn, meta=Label, x expr=\coordindex] {\testdata};
    \addplot [nodes near coords, fill=yellow!60] table [y=Numpy, meta=Label, x expr=\coordindex] {\testdata};
    \addplot [nodes near coords, fill=ForestGreen!60] table  [y=Pandas, meta=Label, x expr=\coordindex] {\testdata};
\legend{Tensorflow, 
PyTorch, 
Sickit-Learn,
Numpy,
Pandas, 
}
\end{axis}
\end{tikzpicture}}
  \caption{Distribution of fixing patterns across different libraries.}
  \label{fixingpatterndist}
\end{figure}
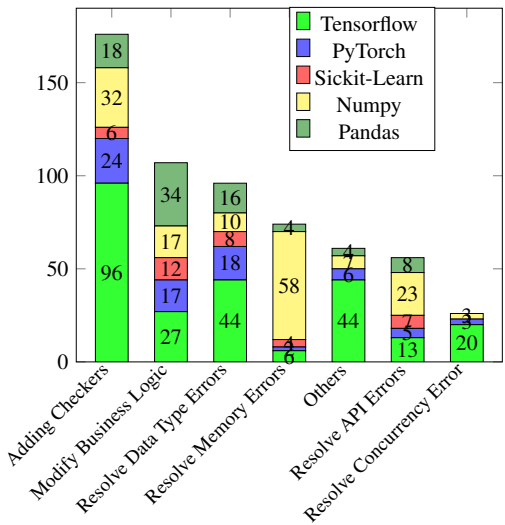

\begin{tcolorbox}[width=8.5cm]
\textbf{Finding 6}: \textbf{Adding Checkers} is the most common fixing pattern across the studied ML libraries accounting for 17.9\% of vulnerabilities.
\end{tcolorbox}

\begin{table}[t!]
\caption{Subcategories of \textbf{Add Checkers}}
\label{checkerstable}
\setlength{\tabcolsep}{14pt} 
\renewcommand{\arraystretch}{1}
\resizebox{\columnwidth}{!}{%
\begin{tabular}{ccccc}
\hline
& ACTP$^1$ & ACFO$^2$ &ACNP$^3$ & ACR$^4$ \\
\hline
Tensorflow   & 67  & 13 & 16 & 0 \\
PyTorch      & 13  & 7  & 3  & 1 \\
Sickit-learn & 1   & 4  & 1  & 0 \\
Pandas       & 7   & 10 & 1  & 0 \\
Numpy        & 14  & 1  & 13 & 4 \\
\hline \hline
Sum          & 102 & 35 & 34 & 5 \\
\hline
\end{tabular}
}
\vskip3pt
{\small
$^1$ Add Checker for Tensors Property $|$ 
$^2$ Add Checker for Overflow $|$
$^3$ Add Checker for Null Pointer Dereference$|$
$^4$ Add Checker for Recursion}
\end{table}

Figure \ref{rootsymp} illustrates the mapping of root causes to their corresponding fixing patterns. As shown in the figure, \textbf{Memory Errors} have three common fixing patterns including \textbf{Improper Memory Management}, and \textbf{Add Checker for Tensor Property}, \textbf{Add Checker for Null Pointer}. The major fixing pattern for \textbf{Memory Errors} is Memory Release Management where developers use memory management APIs to free allocated memories after their effective lifetime. This pattern is mostly used by the Numpy library community which is developed in C language and developers should take care of memory management manually. This \textbf{Memory Leak} vulnerability\footnote{https://github.com/numpy/numpy/commit/\\4e19f408de900f958441af4ec8a458f5ce6473eb} perfectly explains how developers use \textbf{Memory Release Management} pattern to resolve the problem. In this vulnerability, the \textit{slice} object is not decref'd, hence the developer calls \textit{Py\_DECREF()} api with \textit{slice} as the parameter to fix the \textbf{Memory Leak} problem.

The second fixing pattern which resolves \textbf{Memory Errors} is \textbf{Add Checker for Tensor Property}. Often, tensor properties can be problematic if developers do not employ appropriate checkers, either project-specific or general checkers. This vulnerability is a good example of how developers use checkers to prevent \textbf{Infinite Loop}. In this case, \textbf{Infinite Loop} occurs because \textit{num\_cols} exceeds $2^{31-1}$ which is the maximum value an int32 bits variable can take. The developer uses two checkers of \textbf{OP\_REQUIRES} kind to guard against \textbf{Infinite Loop} via checking the dimensions of the input data must be less than or equal to $2^{31-1}$.

Figure \ref{rootsymp} further shows that \textbf{Add Checker for Tensor Property} is the major fixing pattern to fix \textbf{Data Type Errors}. Often lack of checking tensor properties is the root cause of \textbf{Data Type Errors}, e.g. \textbf{Integer Overflow}. In this example\footnote{https://github.com/tensorflow/tensorflow/commit/\\4253f96a58486ffe84b61c0415bb234a4632ee73} which is \textbf{Integer Overflow}, there is no checker on \textit{data[axis]} to make sure it does not exceed in32 bits range limits. The developer overcomes the problem by adding a checker of \textbf{TF\_LITE\_ENSURE} on line 76 of \textit{ tensorflow/lite/kernels/concatenation.cc}.

\begin{tcolorbox}[width=8.5cm]
\textbf{Finding 7}: \textbf{Lack of Checking Tensor Property} is a common cause of \textbf{Data Type Errors} and \textbf{Memory Errors} in the studied ML libraries. Developers can overcome the vulnerabilities by using \textbf{Add Checker for Tensor Property} as the fixing pattern. 
\end{tcolorbox}

\subsection{RQ5: Fixing Effort}

\begin{figure}[t!]
    \centering
      \includegraphics[width=12cm]{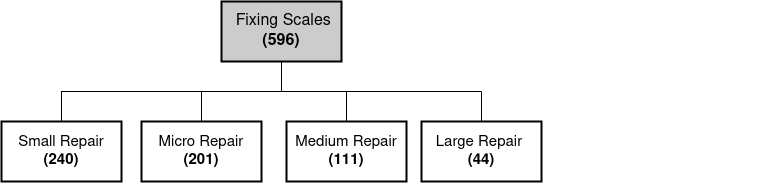}
    \caption{Taxonomy of fixing effort in the studied ML libraries.}
   \label{scalediag}
\end{figure}

\begin{figure}[t!]
     \centering
  \renewcommand{\arraystretch}{1}
\resizebox{0.8\columnwidth}{!}{%
    \begin{tikzpicture}
    \begin{axis}[
        ybar stacked,
    	bar width=15pt,
        ymin=0,
        ymax=217,
        xtick=data,
        legend style={at={(0.6,0.8)},anchor=west},
       xticklabel style={rotate=45,anchor=east},
        xticklabels={\small{Data Type Errors}, \small{Memory Errors}, \small{API Errors}, \small{Business Logic Errors}, \small{Others}, \small{Concurrency Errors}},
    ]
    \addplot [ nodes near coords, fill=green!80] table [y=MicroRepair, meta=Label, x expr=\coordindex] {\testdata};
    \addplot [nodes near coords, fill=blue!60] table   [y=SmallRepair, meta=Label, x expr=\coordindex] {\testdata};
    \addplot [nodes near coords, fill=red!60] table    [y=MediumRepair, meta=Label, x expr=\coordindex] {\testdata};
    \addplot [nodes near coords, fill=yellow!60] table [y=LargeRepair, meta=Label, x expr=\coordindex] {\testdata};
\legend{MicroRepair, 
SmallRepair, 
MediumRepair,
LargeRepair,
}
\end{axis}
\end{tikzpicture}}
\caption{Distribution of fixing effort regarding different root causes.}
 \label{rootvsscale}
\end{figure}
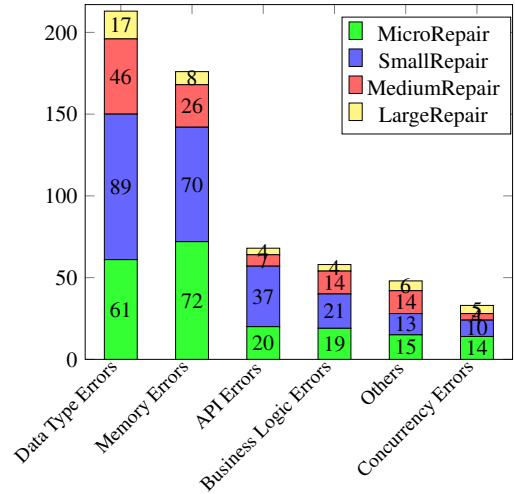

The taxonomy of fixing effort of the studied ML vulnerabilities is shown in Figure \ref{scalediag}, which is organized into four categories. We adopted the categories from \cite{di2017comprehensive, tan2014bug,li2017large} to show the scales of fixing effort: 1) \textbf{Micro repair}: 0-50 added or deleted lines, 2) \textbf{Small repair}: more than 50 added or deleted lines, 3) \textbf{Medium repair}: 50-200 added or deleted lines, and 4) \textbf{Large repair}: more than 200 added or deleted lines. As shown in Figure \ref{scalediag}, 73.9\% of vulnerabilities can be fixed by micro and small fixing efforts. Also, Figure \ref{rootvsscale} further shows the distribution of fixing scales across different root causes. As we can see from the figure, the distributions of fixing effort of vulnerabilities of different root causes do not have dramatic difference, which may suggest the root cause of a vulnerability does not affect its fixing effort. 

\begin{tcolorbox}[width=8.5cm]
\textbf{Finding 8}: Most (73.9\%) vulnerabilities in the studied ML libraries can be fixed in small scales.  Fixing effort of a vulnerability is not related to its root cause.
\end{tcolorbox}

\definecolor{ForestGreen}{RGB}{34,139,34}

\pgfplotstableread{
Label MicroRepair SmallRepair MediumRepair LargeRepair
DataTypeErrors 61 89 46 17
MemoryErrors 72 70 26 8
APIErrors 20 37 7 4
InconsistentBusinessLogic 19 21 14 4
Others 15 13 14 6
ConcurrencyErrors 14 10 4 5
    }\testdata

\section{Implications}

Our study reveals several interesting findings that can serve as practical guidelines for both industry and academic communities to improve software security development for ML libraries.

\noindent \textbf{Avoid Data Type Errors.}
According to our root cause analysis and finding 2 in Section~\ref{section:RQ2}, we conclude that \textbf{Data Type Errors} are the most common root cause of vulnerabilities. As a result, we suggest developers pay attention to 1) the range of integer variables they define, e.g., developers should define int64 bits instead of int32 bits to prevent overflow or integer truncation, 2) developers should pay attention to type conversions as it is one of the main reasons for integer overflow, and 3) developers should use checkers (either library-specific, e.g.,  \textit{OP\_REQUIRES}, or if modules) to make sure that there are no vulnerabilities related to tensor properties, i.e., shapes, ranks, values, or elements.

\noindent \textbf{Always Initialize Variables.} Our analysis shows that 7.8 \% of vulnerabilities are due to initialization issues, e.g., lack of initializing tensors or variables. It is always better to initialize variables or any resource during the development of ML tasks.

\noindent \textbf{Apply Dynamic Analysis Tools to Avoid Memory Related Vulnerabilities.}
We find that \textbf{Resource Consumption} is the fourth most common symptom of vulnerabilities accounting for 19.4 of vulnerabilities. Often, understanding whether libraries are suffering from \textbf{Resource Consumption} is challenging since test suits can not profile resource usage. Moreover, we find that \textbf{Memory Errors} is the second most root cause of vulnerabilities in ML libraries. The main reason for \textbf{Memory Errors} is memory release management, where developers forget to release the allocated memories. Hence, we strongly suggest developers of studied ML libraries use dynamic checkers to detect and avoid memory-related vulnerabilities by applying memory checkers, e.g., Valgrind\footnote{https://valgrind.org/}. We have manually checked that Valgrind can detect most half of the \textbf{Memory Errors} related vulnerabilities, which suggests the importance of using dynamic analysis tools to avoid memory-related vulnerabilities.

\noindent \textbf{Use API Usage Checker.} As shown in Figure~\ref{rootcausediag}, \textbf{API Errors} is also a common root cause for vulnerabilities in ML libraries; API-related issues mainly cause vulnerabilities whose root causes fall in this category during the development of ML libraries. This root cause is decomposed into different subcategories as shown in Table~\ref{apierror}. As we can see, \textit{API Misuse}, \textit{Using Wrong API}, and \textit{API Version Issue} cover 50 out of the 68 \textbf{API Errors} involved vulnerabilities. We have checked that most of these types of API usage issues can be avoided by using API usage checkers~\cite{zhang2020python,zhang2021unveiling}, thus developers are suggested to apply these API misuse detectors when working on ML development tasks for avoiding \textbf{API Errors} related vulnerabilities. The third most common API inconsistency is the API Version Issue, where developers mostly use improper versions of APIs, causing different vulnerabilities. Hence, there is a need to develop version control (checking) tools to assist developers in using up-to-date APIs. 

\section{Actionable Applications of Our Findings}
\label{sec:tool}
\begin{figure}[t!]
	\centering
  \renewcommand{\arraystretch}{4}
\resizebox{\columnwidth}{!}{%
\includegraphics[width=\textwidth]{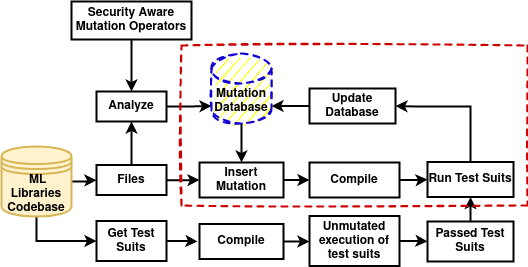}}
	\caption{Architecture of {\tool}.}
	\label{deepmutdiag}
\end{figure}

We believe the findings of our study can be used to improve software vulnerability development tasks, such as detecting similar unknown vulnerabilities based on the studied vulnerabilities in this work, categorizing newly reported vulnerabilities, and improving security vulnerability testing via security mutation analysis.

To make our finding actionable, we take security mutation analysis as an example and develop {\tool}, an automated mutation testing tool for ML libraries, as a proof-of-concept application of our findings. {\tool} is designed to evaluate the adequacy of the existing test suite of ML libraries against security-aware mutation operators extracted from the vulnerabilities studied in this work. Specifically,  we use all the \textbf{Tensor property Issues} related vulnerabilities from TensorFlow as an example to extract mutation operators for {\tool}.

Figure \ref{deepmutdiag} shows the overall architecture of {\tool}, which has the following steps to find alive mutants. First, it gets all source files from the target ML library then performs an initial analysis to extract potential statements that match with the extracted mutation operators. 
The potential statements are stored in the mutation database. The mutation loop starts iterating over the database and inserts each mutant into the target source code. Once the insertion is finished, the compilation process begins where the library under test is compiled. Subsequently, all test suites run against inserted mutants. To determine whether the mutant is killed or not, we perform a simple analysis on the stack trace of running test suites and update the database. This process continues until all mutants in the database are executed. 

We performed {\tool} on \textit{TensorFlow/core/kernel module} module. We ran {\tool} on branch v2.7.0\footnote{https://github.com/TensorFlow/TensorFlow/tree/r2.7}, one of the active branches of TensorFlow. It took around one week to get the initial results. In total, {\tool} generates more than 3k mutants, and among them 1.2K are alive.

\begin{figure}[t!]
	\centering
  \renewcommand{\arraystretch}{6}
\resizebox{\columnwidth}{!}{%
    \includegraphics[width=9cm]{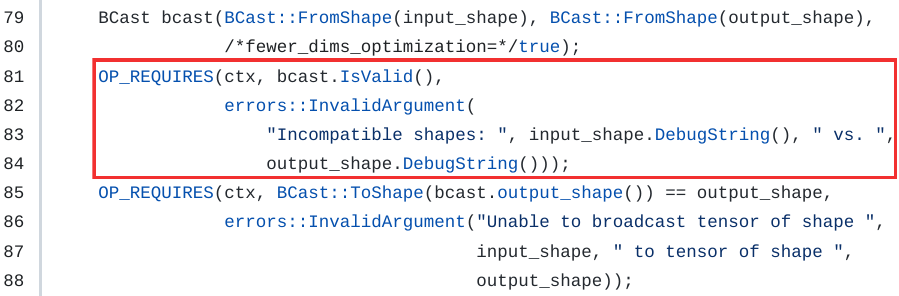}}
	\caption{An example of alive mutant found in Tensorflow kernel module.}
	\label{alivem}
\end{figure}

Here, we further show an example of an alive mutant found by {\tool} from \textit{broadcast\_to\_ops.cc} file (from TensorFlow kernel module) illustrated in Figure \ref{alivem} that is not covered by the TensorFlow kernel's test suite. As you can see, a checker at lines 81-84 is responsible for checking whether input and output shapes are compatible with each other. If the shapes violate the constraints defined in the checker, a run time error will be generated. First, {\tool} analyzes \textit{broadcast\_to\_ops.cc} to search for potential mutation locations in this file and applicable mutation operators, as a result, 
{\tool} 
finds an applicable operator (i.e., eliminating \textit{OP\_REQUIRES} and \textit{OP\_REQUIRES\_OK} from source code) for the fixing patch of existing vulnerabilities in TensorFlow. To apply the operator, {\tool} removes lines 81-84 and compiles TensorFlow. Once the compilation is done, {\tool} runs all test suites of kernel modules and finds that none of the existing tests can detect the deleted lines. 


We have reported these live mutants to the security team of TensorFlow. They confirmed that these scenarios are not covered by the existing test suite and agree that {\tool} is useful to help design security test cases. In this paper, we apply {\tool} on the TensorFlow core kernel module. However, {\tool} is extensible to more components or more ML libraries.

\section{Threats to validity}
\label{sec:threats}

\paragraph{Internal Validity.}
The main internal threat to our work is our manual analysis labeling and classification of software security vulnerabilities which may suffer from subjective bias and errors. To guard against this, two Ph.D. students have reviewed the collected commits in multiple rounds. The students discuss any possible disagreement after each round until a consensus is reached.

\paragraph{External Validity.}
The dominant threat to the external validity of this study is the collected dataset. To overcome this threat, we collected commits from five different ML libraries; two are very famous and widely used DL libraries, including TensorFlow and PyTorch; one of them is Scikit-learn which is a renowned classical ML library which often is used beside DL libraries. We also collected data from two well-known data analytics and visualization tools, including Pandas and Numpy. The reason behind this diverse data collection is to generalize our findings to wide domains and increase the reliability of findings. To augment our classification process and make them more accurate, besides reviewed commits, we also reviewed issues and merged pull requests linked to the parent commits. Please see Section \ref{datacoll} for further details.
\section{Related Work}

\subsection{Studies on General Vulnerabilities}
\label{sec:sbc}

This section surveys the existing studies on analyzing the characteristics of vulnerabilities in general software projects. 

\subsubsection{Vulnerabilities in General Software Systems}
There are many efforts to characterize software security vulnerabilities in traditional software systems \cite{bosu2014identifying,jimenez2016empirical,tan2014bug}.

Jimenez et al.~\cite{jimenez2016empirical} analyzed characteristics of vulnerabilities of Linux kernel and OpenSSL. They collected 2k vulnerable git commits that are 1) reported in CVE, 2) have vulnerable keywords, and 3) have a CVE number in their message and title. They find 20 frequent types of vulnerabilities; among them, CWE-200 and CWE-119 are the dominant ones for Linux and CWE-119, CWE-399, and CWE-362, for OpenSSL. There are two significant differences with our analysis; 1) Based on their findings,  \textbf{Numeric} and \textbf{Memory} are not major common vulnerabilities in general software systems, 2) Unlike studied ML libraries, general software systems require complex efforts to fix vulnerabilities.

Tan et al. \cite{tan2014bug} conducted an empirical study on three notable projects, including Linux kernel, Mozilla, and Apache, via analyzing around 2k real-world bugs. They revealed that semantic bugs are the major common bugs in general software systems, and memory bugs decrease as they evolve. The significant difference with our analysis is that they did not introduce vulnerability types; instead, they focused on root causes analysis. Also, their analysis is based on general and vulnerable related bugs, while we do not cover the general bugs. Bosu et al.~\cite{bosu2014identifying} analyzed code review requests from 10 software projects to identify vulnerable code changes. They developed a tool called \textit{Gerrit-Miner} that mines code reviews from the Gerrit code review portal that is publicly available. They mined more than 260k reviews and analyzed 1k reviews thoroughly for the analysis. They find that \textit{Race Condition} and \textit{Buffer Overflow} are the most common vulnerability types in traditional software systems. These findings are not aligned with our work where \textbf{Race Condition} and \textbf{Buffer Overflow} are the least common vulnerability types.

\subsubsection{Vulnerabilities in Software Ecosystems}
Software ecosystems are vital in modern software development as they provide reusable packages to developers and increase development speed. Two notable ecosystems are \textit{npm}\footnote{https://www.npmjs.com/} that supports \textit{Node.js} packages and \textit{PyPi}\footnote{https://pypi.org/} that supports Python packages. Alfadel et al.~\cite{alfadel2021empirical} conducted a study to characterise vulnerabilities in \textit{PyPi}. They focused mostly on how long it takes to find and fix vulnerabilities in projects in  \textit{PyPi}.
They found that  \textit{Cross-Site-Scripting (XSS)} and \textit{Denial of Service (DoS)} are the foremost common vulnerability types in \textit{PyPi}, which are significantly different from the common vulnerability types in ML libraries studied in this paper. Zimmerman et al.~\cite{zimmermann2019small} conducted an empirical study on \textit{npm} ecosystem to analyze the dependencies among public users of packages, their maintainers, and corresponding public security reports.  
They find that a single point of failure is the primary vulnerability of \textit{npm} because \textit{npm} packages are often not maintained constantly, which makes large codebases vulnerable. The significant difference with the studied ML libraries is the mitigation strategies where for example, in \textit{npm}, trusted maintainers and code vetting process are two promising fixing strategies.

\subsubsection{Vulnerabilities in Android applications}
\label{sec:androidVul}
There exist many studies on the vulnerabilities of Android applications~\cite{mazuera2019android,linares2017empirical, jimenez2016profiling}. 

Mazura \cite{mazuera2019android} conducted a large-scale empirical study on Android vulnerabilities by analyzing more than 1k cases from different aspects, including the type of vulnerabilities, the evolution of vulnerabilities, CVSS vendors, the impact of vulnerabilities, and whether they have survived in Android history or not. They find that the significant vulnerability in Android applications is {Permissions, Privileges, and Access Controls}. Linares et al.~\cite{linares2017empirical} characterized different types of vulnerability that may affect android apps, the affected subsystems, and the time it takes to fix vulnerabilities. Similar to \cite{mazuera2019android}, they also mined vulnerabilities from the Android Security Bulletins and the CVE portal. They find that \textit{Memory} and \textit{Data} are the significant types of vulnerability in Android applications. Jimenez et al.~\cite{jimenez2016profiling} performed an empirical study to analyze vulnerabilities of Android applications reported in the National Vulnerability Database. 
They found that 
\textit{Missing/incorrect implementation of features} is the dominating vulnerability type. 
Different from the above studies, we explore vulnerabilities in ML libraries in this paper.

\subsection{Studies on ML Bugs}
\subsubsection{Studies on ML API Usage Bugs}
Islam et al. \cite{islam2019comprehensive} conducted the first empirical study on API usage bugs of five DL libraries, including Caffe, Keras, TensorFlow, Theano, and Torch.
They collected data from Stackoverflow posts, and Github commits to perform their manual analysis. The authors analyzed bug types, root causes, and impact of bugs in DL libraries and found that data and logic-related bugs are the most common bugs in DL libraries. Zhang et al.~\cite{zhang2018empirical} studied DL application bugs built on top of TensorFlow and collected bugs from both Stackoverflow and Github projects. They find that fixing patterns and root causes correlate and suggest developers and researchers make automated bug detection approaches on top of root causes. Humbatova et al.~\cite{humbatova2020taxonomy} provided an extensive and comprehensive taxonomy of faults in DL libraries. They focused on TensorFlow, Keras, and PyTorch for their study. The notable difference of their work with existing studies is that they interviewed 20 researchers and practitioners to increase the reliability of their findings. There are a couple of differences with our work. First, our study merely focuses on Github commits while their study also mined data from Stackoverflow posts. Second, they analyzed general bugs of DL libraries while we studied security vulnerabilities reported in CWE and CVE portals.

\subsubsection{Studies on ML Implementation Bugs}

Thung et al.~\cite{thung2012empirical} studied data from three popular java-based ML libraries to characterize bugs related to the implementation of such tools. Such data are linked to bug reports and bug repositories of the subject programs extracted from the JIRA issue tracking system. Consequently, they came up with 500 bugs and addressed the research questions. They find that algorithmic relayed bugs are the most prevalent in the studied ML libraries. Jia et al.~\cite{jia2021symptoms} conducted an empirical study on implementation bugs of TensorFlow. More specifically, they targeted more than 36k Github projects that use TensorFlow and extract pull requests, bug reports, and code changes from the corresponding repositories to address the research questions. The significant finding of their work is that root causes and symptoms of bugs in TensorFlow are similar to traditional software systems.
The most related papers to our study are the studies conducted by Franco et al.~\cite{di2017comprehensive} and Shen et al.~\cite{shen2021comprehensive}. Franco et al. \cite{di2017comprehensive} conducted the first study on characteristics of real-world numerical bugs of different numerical libraries, including NumPy, SciPy, LAPACK, GNU Scientific Library, and Elementa. They find that 32\% of bugs in the studied libraries are related to \textbf{Numeric}. Our study complements their analysis in the sense that ours is more general since we study both numerical and ML libraries. Also, our analysis is more comprehensive because, besides \textbf{Numeric}, we introduce multiple significant vulnerabilities that are common in numerical and ML libraries. Shen et al.~\cite{shen2021comprehensive} proposed a comprehensive study on DL compiler bugs by manually analyzing 596 bugs from TVM from Apache, Glow from Facebook, and nGraph from Intel.  
They find that type-related bugs are the foremost common bugs in DL compilers. Despite these efforts, the characteristics of software security vulnerabilities have not been well studied, which is the main contribution of this work.

\section{Conclusion}

This paper conducts the first empirical study to understand the characteristics of software security vulnerabilities of ML libraries. The primary motivation behind this study is to help developers of such libraries design and develop vulnerability detection and debugging techniques to increase their quality and reliability. To achieve this goal, we manually analyzed 596 commits from five widely used ML libraries, including TensorFlow, PyTorch, Scikit-Learn, Pandas, and Numpy. The outcome of this study is 19 vulnerability types, 18 root causes, 5 symptoms, 22 fixing patterns, 4 fixing scales, and ultimately 8 findings. Based on these findings, we further provide a set of actionable guidelines to developers and the community to design and develop software vulnerability detection and debugging techniques to increase ML libraries' security.



\section*{Availability}
We make the dataset and source code of our experiments available at  \url{https://cse19922021.github.io/Deep-Learning-Security-Vulnerabilities/}. 



\bibliographystyle{plain}
\bibliography{paper}

\end{document}